\documentclass[11pt,letterpaper]{article}

\usepackage{jcappub}
\usepackage{amsfonts}
\usepackage{amsmath}
\usepackage{amssymb}
\usepackage{aas_macros}
\usepackage{graphicx}
\usepackage{color}
\usepackage{float}
\usepackage{hyperref}

\def\lsim{~\rlap{$<$}{\lower 1.0ex\hbox{$\sim$}}}
\def\bsim{~\rlap{$>$}{\lower 1.0ex\hbox{$\sim$}}}

\def\hmpc{\ {\rm {\it h}^{-1}Mpc}}

\def\apjl{Astrophys.\ J.\ Lett.}

\def\aap{Astron.\ Astrophys.}

\def\apjs{Astrophys.\ J. Supp.}

\def\la{\langle}
\def\ra{\rangle}

\let\ln\relax
\DeclareMathOperator{\ln}{ln}
\DeclareMathOperator{\tr}{tr}

\def\mathbi#1{\textbf{\em #1}}

\newcommand{\bm}[1]{{\mbox{\boldmath $#1$}}}

\newcommand{\J}{\mathcal{J}}
\newcommand{\K}{\mathcal{K}}
\newcommand{\Q}{\mathcal{Q}}

\newcommand{\plin}{P_\text{L}}

\newcommand{\dof}{D}

\def\veta{{\boldsymbol\eta}}

\def\kvh{\hat{\mathbi{k}}}

\def\vk{\mathbi{k}}
\def\vq{\mathbi{q}}

\def\vw{\mathbi{w}}
\def\vx{\mathbi{x}}
\def\vy{\mathbi{y}}

\def\npk{n_\text{pk}}

\def\bnpk{\bar{n}_\text{pk}}

\def\dpk{\delta_{\rm pk}}

\newcommand{\fnl}{f_\text{NL}}

\renewcommand{\v}[1]{\bm{#1}}
\def\d{\delta}
\def\refeq#1{Eq.~(\ref{eq:#1})}
\def\refeqs#1#2{Eq.~(\ref{eq:#1})--(\ref{eq:#2})}
\def\xfl{\v{x}_\text{fl}}
\newcommand{\be}{\begin{equation}}
\newcommand{\ee}{\end{equation}}

\def\comment#1{}
\definecolor{RedWine}{rgb}{0.743,0,0}
\definecolor{RoyalBlue}{rgb}{0.25,.41,.88}
\definecolor{ForestGreen}{rgb}{.13,.54,.13}
\definecolor{DeepPurple}{rgb}{.72,.18,1}

\title{Tidal shear and the consistency of microscopic Lagrangian halo approaches}

\author[a]{Vincent Desjacques,}
\author[b]{Donghui Jeong,}
\author[c]{and Fabian~Schmidt}

\affiliation[a]{Physics department and Asher Space Science Institute, Technion, 3200003 Haifa, Israel}
\affiliation[b]{Department of Astronomy and Astrophysics, and Institute for Gravitation and the Cosmos, 
The Pennsylvania State University, University Park, PA 16802, USA}
\affiliation[c]{Max-Planck-Institut f\"ur Astrophysik, Karl-Schwarzschild-Stra\ss e~1, 85748 Garching, Germany}

\emailAdd{dvince@physics.technion.ac.il}
\emailAdd{djeong@psu.edu}
\emailAdd{fabians@mpa-garching.mpg.de}

\date{\today}

\label{firstpage}

\abstract{%
We delineate the conditions under which the consistency relation for the 
non-Gaussian bias and the universality of the halo mass function hold in the 
context of microscopic Lagrangian descriptions of halos. 
The former is valid provided that 
the collapse barrier depends only on the 
physical fields (instead of fields 
normalized by their variance for example) and explicitly includes 
the effect of {\it all} physical fields such as the tidal shear. 
The latter holds provided that the response of the halo number density to a 
long-wavelength density fluctuation is equivalent to the response induced by 
shifting 
the spherical collapse threshold.
Our results apply to any Lagrangian halo bias prescription. 
Effective ``moving'' barriers, which are ubiquitous in the literature, do not 
generally satisfy the consistency relation. Microscopic barriers including the 
tidal shear lead to two additional, second-order Lagrangian bias parameters 
which ensure that the consistency relation is satisfied. 
We provide analytic expressions for them.
}

\begin{document}

\maketitle

\section{Introduction}

The analysis of ongoing and future galaxy redshift surveys will rely heavily on galaxy clustering, which
has a very large signal-to-noise and correspondingly information content
\cite{eBOSS,HETDEX,DESI,PFS,amendola/etal:2013,WFIRST}.
However, our ability to interpret the observational data is limited because the distribution of observed large scale
structure tracers, such as galaxies, is biased relative to that of the matter (see \cite{Desjacques:2016bnm} for a
recent comprehensive review). 
Crucially, the bias parameters are not all free but satisfy certain \emph{consistency relations} which can be used to
constrain or measure the biases.
For example, the Lagrangian LIMD (local in matter density) bias parameters are directly related to the response of the
mean comoving density of galaxies $\bar{n}_g$ under a change of the background comoving matter density $\bar\rho_m$
of the universe \cite{Kaiser:1984sw,Cole:1989vx,Mo:1995cs,Sheth:1999mn,baldauf/seljak/senatore:2011,PBSpaper},
\be
b_N^L = \frac{\bar\rho_m^N}{\bar{n}_g} \frac{\partial^N \bar{n}_g}{\partial \bar\rho_m^N}\;,
\label{eq:PBS0}
\ee
which is the essence of the \emph{peak-background split} (PBS) \cite{Bardeen:1985tr}.
This identity has been used to measure the bias parameters of dark matter halos in ``separate universe'' simulations
\cite{lazeyras/etal,baldauf/etal:2015,li/hu/takada:2016}.
Bias also leads to a pronounced, large-scale signature in the clustering of biased tracers for certain types of inflationary
models \cite{Dalal:2007cu}.
Specifically, in the presence of local-type primordial non-Gaussianity, the large-scale bias relation is given by
\be
\d_g(\vx,\tau) = b_1(\tau) \d(\vx,\tau) + 2\fnl b_\phi(\tau) \phi(\vx)\,,
\ee
where $\phi$ is the primordial Bardeen potential, and $\fnl$ is the dimensionless amplitude of primordial non-Gaussianity.
The coefficient $b_\phi$ of this non-Gaussian bias also follows from a peak-background split relation
\cite{slosar/hirata/etal:2008}:
\begin{equation}
\label{eq:PBS}
b_\phi = 2 \frac{\partial\ln\bar n_h}{\partial\ln{\cal A}_s} = \frac{\partial\ln\bar n_h}{\partial\ln\sigma_8} \;,
\end{equation}
where ${\cal A}_s$ is the root-mean-squared (r.m.s.) amplitude of the linear power spectrum and $\sigma_8$ is the 
r.m.s. variance of linear density fluctuations smoothed on a comoving scale of 
$8\hmpc$.
Eq.~(\ref{eq:PBS}) follows from the local effect which a long-wavelength potential perturbation induces in the presence of local-type
primordial non-Gaussianity, and has to be satisfied for any physical tracer (just as \refeq{PBS0}); more general relations can
be obtained for other types of primordial non-Gaussianity
\cite{schmidt/kamionkowski:2010,long,PBSpaper,assassi/baumann/schmidt}.
Not surprisingly, numerical analyses have confirmed the validity of this non-Gaussian bias consistency relation for halos
\cite{Dalal:2007cu,Desjacques:2008vf,Grossi:2009an,short,smith/ferraro/loverde:2012}, 
also when the halo mass function is not universal, as is the case for spherical overdensity (hereafter SO) halos
\cite{baldauf/etal:2015,biagetti/lazeyras/etal:2016}. 

In the literature, the halo mass function (the number of halos per unit comoving volume and logarithmic mass interval)
\be
\bar{n}_h(M) = \frac{\bar\rho_m}{M}\nu_c f(\nu_c)\frac{d\ln \nu_c}{d\ln M}
\label{eq:nhnufnu}
\ee
is dubbed universal if the multiplicity function $f$ is a function of the peak height or significance $\nu_c\equiv \d_c/\sigma_0$
only. Here, $\sigma_0$ is the r.m.s. value of the linear density field smoothed at the halo mass scale $M$, and $\d_c$ is the
critical density for spherical collapse (e.g., \cite{Gunn:1972sv}).
If this is the case, then the same multiplicity function $f(\nu_c)$ can be used to predict halo abundance for different cosmologies
and redshifts. 
For example, Press-Schechter and excursion set theory \cite{press/schechter:1974,Bond:1990iw} which, in their original formulation,
rely on the statistics of the linear density field only, both yield universal mass functions.
Another famous example of universal mass function is the Sheth-Tormen fitting formula \cite{sheth/mo/tormen:2001}.
By contrast, Ref.~\cite{tinker/kravtsov/etal:2008} showed that their mass function is not universal (although they used a multiplicity
function of the form $f(\sigma_0^{-1})$, essentially equivalent to $f(\nu_c)$, but with redshift-dependent coefficients).

However, as emphasized in \cite{despali/giocoli/etal:2015}, the amount of non-universality reported crucially depends on the halo
definition, unlike the bias consistency relations which hold regardless of the halo definition.
Furthermore, in the context of Lagrangian models of halo formation, this definition of universality is too restrictive as it excludes
for example the peak multiplicity function, which depends also on moments of the density power spectrum
\cite{Bardeen:1985tr,Paranjape:2012jt}.
Therefore, the seemingly less restrictive condition
\begin{equation}
  \label{eq:universal}
  \frac{\partial\ln\bar n_h}{\partial\ln\sigma_8} = b_1^L \d_c \;,
\end{equation}
where $b_1^L=b_1 -1 $ is the halo Lagrangian LIMD bias, appears more appropriate mainly for two reasons:
i) it includes multiplicity function {\it \`{a} la} peak theory; and 
ii) it makes the connection between universality and bias explicit.

In this paper, we investigate these issues in the framework of \emph{microscopic Lagrangian descriptions of halos.} 
Consider the fractional overdensity of halos identified at time $\tau_0$, $\d_h(\vx,\tau_0)$.
We define the Lagrangian halo density contrast $\d_h^L(\vq,\tau_0)$ through
\begin{equation}
\d_h^L(\vq,\tau_0)
= \frac{1 + \d_h[\xfl(\vq,\tau_0),\tau_0]}{1 + \d[\xfl(\vq,\tau_0),\tau_0]} - 1 \;,
\end{equation}
where $\xfl(\vq,\tau)$ denotes the fluid trajectory for a Lagrangian coordinate $\vq$, and $\d(\vx,\tau)$ denotes the matter density perturbation. 
A microscopic Lagrangian description of halos is defined as an explicit functional relation
\begin{equation}
  \d_h^L(\vq,\tau_0) = \mathcal{F}[\d^{(1)}(\vq',\tau_0), \tau_0]\;,
\end{equation}
Here, $\mathcal{F}$ is a spatial functional of the linear matter density field. 
In the most commonly considered cases, this functional becomes a local function of the linearly extrapolated density field
$\d^{(1)}_R(\vq)\equiv\int {\rm d}^3\vy\, W_R(y)\d^{(1)}(|\vq-\vy|)$ filtered on a scale $R$, and a finite set of its spatial derivatives:
\begin{equation}
  \d_h^L(\vq,\tau_0) = F\Big(\d^{(1)}_R(\vq,\tau_0),\  \partial_i \d^{(1)}_R(\vq,\tau_0),\  \partial_j\partial_k \d^{(1)}_R(\vq,\tau_0),\  \tau_0\Big)\,.
  \label{eq:dhmicro}
\end{equation}
The most important examples of such a prescription include peaks of the density field \cite{Bardeen:1985tr},
which can be unified with excursion set theory into the so-called excursion set peaks (ESP) approach \cite{Paranjape:2012ks,Paranjape:2012jt}.
General expressions for the Lagrangian halo biases derived from this type of description have been obtained in, e.g., \cite{Lazeyras:2015giz}
under the spherical-collapse assumption. 
In particular, the Lagrangian LIMD, tidal shear and higher-derivative bias parameters satisfy a hierarchy of consistency relations which can be used
to extract information on these Lagrangian bias parameters 
\cite{Musso:2012ch,Paranjape:2012jt,Biagetti:2013hfa,Modi:2016dah,castorina/etal:2017,Castorina:2016tuc,Chan:2017azj}.
Here, we will outline how the peak and ESP approaches can be generalized to include the effect of the tidal shear on the gravitational collapse of
the halo, leading to
\begin{equation}
  \d_h^L(\vq,\tau_0) =
  F\Big(\partial_l\partial_m \Phi^{(1)}_R(\vq,\tau_0),\  \partial_i \d^{(1)}_R(\vq,\tau_0),\  \partial_j\partial_k \d^{(1)}_R(\vq,\tau_0),\  \tau_0\Big)\,,
  \label{eq:dhmicroK}
\end{equation}
where $\Phi^{(1)}_R$ is the smoothed linear gravitational potential. 
The tidal shear has been shown to play a crucial role in the formation of virialized structures, leading to ellipsoidal collapse etc.
\cite{lyndenbell:1964,lin/mestel/etal:1965,Zeldovich:1969sb,hoffman:1986,Bond:1993we,sheth/mo/tormen:2001}.
In several implementations of ellipsoidal collapse, however, the physical barrier $B$ is approximated by a ``moving'' barrier with an explicit
dependence on $\sigma_8$ \cite{sheth/mo/tormen:2001,Zhang:2005ar,Moreno:2007aw,Maggiore:2009rw,Paranjape:2012ks,Sheth:2012fc,Lapi:2013bxa}.
As we will demonstrate, such an approximation leads to the violation of the consistency relation \refeq{PBS}.

Although our calculations rely on a specific ESP model for illustrative purposes, the relations we derive and the conclusions we draw in this paper
are completely general 
in the context of microscopic Lagrangian descriptions of halos [\refeq{dhmicroK}].
For clarity, we shall briefly summarize our main findings here.\\

\textbf{Summary of our key results:} 
\begin{enumerate}
\item The consistency relation \refeq{PBS}, which must be obeyed by any physical tracer, is satisfied in microscopic halo approaches if, and
  only if, the function $F$ in \refeqs{dhmicro}{dhmicroK} does not depend on the amplitude ${\cal A}_s$ of the linear power spectrum, or,
  equivalently, the r.m.s. variance $\sigma_8$ of density fluctuations (see \S\ref{sec:NGconsistency} for details).
\item The scale-dependent bias amplitude $b_\phi$ and the linear Lagrangian bias $b_1^L=b_1-1$ are related by \refeq{universal}
  if, and only if, the response of the
  function $F$ to a long-wavelength density perturbation $\d_\ell^{(1)}$ is
    \begin{equation}
      \label{eq:dfdc}
    \frac{\partial F}{\partial\d_\ell^{(1)}} = - \frac{\partial F}{\partial\d_c} \;.
  \end{equation}
  This requirement stems from a condition on the collapse barrier (see \S\ref{sec:universality} for details).
\end{enumerate}
The paper is organized as follows. In \S\ref{sec:shear}, we generalize the excursion set approach to the tidal shear and
calculate the second-order, tidal shear Lagrangian bias parameters. In \S\ref{sec:bias}, we delineate under which
circumstances i) the non-Gaussian bias consistency relation is violated and ii) the predicted halo mass function is
universal. We conclude in \S\ref{sec:conclusions}.

\section{Shear and non-Gaussian bias in the ESP approach}
\label{sec:shear}

We begin with a derivation of the second-order Lagrangian bias function including the gravitational tidal shear, and provide
expressions for the corresponding Lagrangian shear bias parameters. These results will be useful for the discussion in
\S\ref{sec:bias}. Note that, hereafter, we will denote Lagrangian coordinates with $\vx$.

\subsection{Basic definitions}
\label{sec:basic}

As Lagrangian description of halo clustering, we consider the excursion set peak (ESP) approach \cite{Paranjape:2012ks,Paranjape:2012jt},
which combines both peaks theory \cite{Bardeen:1985tr} and excursion set theory \cite{Bond:1990iw}.
In the ESP framework, each local maximum of the linear density field smoothed on scale $R$ corresponds to a proto-halo of mass
$M\propto R^3$ unless it is embedded in a bigger halo.
This last condition is commonly referred to as the first-crossing condition.
For simplicity, we shall assume a Gaussian filter throughout this paper, so that the slope $d\d_R^{(1)}/dR$ of the
smoothed random walk $\d_R^{(1)}(\vx,\tau)$ with respect to (w.r.t.) the filtering scale $R$ is directly proportional to the
peak curvature $\J_1$ (see Eq.~(\ref{eq:defs1})).
Consequently, the first-crossing condition is equivalent to a positive peak curvature, which is always satisfied for local
density maxima.
Notwithstanding, all our results can be straightforwardly generalized to arbitrary filter functions upon treating the slope
$d\d_R^{(1)}/dR$ as an additional, independent variable \cite[see, e.g.,][]{Desjacques:2013qx}.

We introduce the components $\nu$, $\K_{ij}$, $\eta_i$ and $\zeta_{ij}$ of the smoothed density, shear, density gradient and Hessian,
respectively, all with variance normalized to unity,
\begin{gather}
  \K_{ij}(\tau,\vx)=\frac{1}{\sigma_0(\tau)}\partial_i\partial_j\Phi_R^{(1)}(\tau,\vx)\;, \qquad
  \nu(\tau,\vx) = \tr \K_{ij} = \frac{1}{\sigma_0(\tau)}\d_R^{(1)}(\tau,\vx) \;, \nonumber \\
  \eta_i(\tau,\vx) = \frac{1}{\sigma_1(\tau)}\partial_i\d_R^{(1)}(\tau,\vx) \;, \nonumber \\
  \zeta_{ij}(\tau,\vx)=\frac{1}{\sigma_2(\tau)}\partial_i\partial_j\d_R^{(1)}(\tau,\vx)\;, \qquad
  \J_1(\tau,\vx) = -\tr \zeta_{ij} = -\frac{1}{\sigma_2(\tau)}\nabla^2\d_R^{(1)}(\tau,\vx) \;.
  \label{eq:defs1}
\end{gather}
To avoid clutter, we will hereafter not make the smoothing of the fields explicit.
Furthermore, we will assume that all the fields are evaluated at the epoch of halo collapse, corresponding to a time
$\tau=\tau_0$. Therefore, we shall omit the explicit time dependence in all subsequent expressions.
The gravitational potential $\Phi_R^{(1)}(\vx)$ and the density $\delta_R^{(1)}(\vx)$ are related through the Poisson equation, and
$\sigma_n$ are the spectral moments of the density field smoothed on the Lagrangian halo scale $R$ with a spherically symmetric filtering
function $W_R(k)$,
\begin{equation}
  \sigma_n^2 = \int_{\vk} k^{2n} \,W_R(k)^2\,\plin(k)
\equiv  \int \frac{{\rm d}^3\vk}{(2\pi)^3} k^{2n} \,W_R(k)^2\,\plin(k)\;,
\end{equation}
where $\plin(k)$ is the power spectrum of the density field $\d^{(1)}$ linearly extrapolated to the collapse epoch.
In particular, $\sigma_0$ is the variance of the smoothed density field.
Finally, the convention $\J_1 \equiv - \tr \zeta_{ij}$ and $\nu\equiv + \tr \K_{ij}$ follows historical developments.

We define the traceless parts $\bar\K_{ij}$ and $\bar \zeta_{ij}$ through the relations 
\begin{align}
\bar\K_{ij} &= \K_{ij} - \frac{1}{3}\delta_{ij} \nu \nonumber \\
\bar\zeta_{ij} &= \zeta_{ij} + \frac{1}{3} \delta_{ij} \J_1 \;.
\end{align}
Note that a local density maximum satisfies $\J_1 > 0$.
In Fourier space, these variables become
\begin{align}
\bar\K_{ij}(\vk) &= \frac{1}{\sigma_0}\left(\frac{k_i k_j}{k^2}-\frac{1}{3}\delta_{ij}\right)\d^{(1)}(\vk) \nonumber \\
\bar\zeta_{ij}(\vk) &= -\frac{1}{\sigma_2}\left(k_i k_j - \frac{k^2}{3}\delta_{ij}\right)\d^{(1)}(\vk) \;.
\end{align}
The covariance structure of these variables is well known \cite{doroshkevich:1970,Bardeen:1985tr,vandeWeygaert:1995pz}.
The 1-point auto- and cross-correlation of the components $\bar\K_{ij}$ and $\bar \zeta_{ij}$ are given by
\begin{gather}
\big\la \bar\zeta_{ij}\bar \zeta_{lm}\big\ra =
\big\la \bar\K_{ij}\bar\K_{lm}\big\ra =
\frac{1}{15} \left(\delta_{il}\delta_{jm}+\delta_{im}\delta_{jl}-\frac{2}{3}\delta_{ij}\delta_{lm}\right) \nonumber \\
\big\la \bar\zeta_{ij}\bar\K_{lm}\big\ra = -\frac{\gamma_1}{15} 
\left(\delta_{il}\delta_{jm}+\delta_{im}\delta_{jl}-\frac{2}{3}\delta_{ij}\delta_{lm}\right) \;.
\end{gather}
The dimensionless coefficient $\gamma_1=\sigma_1^2/(\sigma_0\sigma_2)$, which varies in the range $0<\gamma_1<1$, measures the shape of
the smoothed power spectrum $W_R^2\plin(k)$ \cite{Bardeen:1985tr}.
It is convenient to introduce the rotational invariants
\begin{equation}
\K_2 \equiv \frac{3}{2}\tr\big(\bar\K_{ij}^2\big) \;,\qquad
\J_2 \equiv \frac{3}{2}\tr\big(\bar \zeta_{ij}^2\big) \;,\qquad
\Q_2 \equiv \frac{3}{2}\tr\big(\bar\K_{ij}\bar\zeta_{ij}\big) \;.
\end{equation}
With this, we have
\begin{equation}
\la \K_2 \ra = \la \J_2 \ra = 1\,,\quad \la \K_2 \J_2 \ra = - \gamma_1\;.
\end{equation}
It can be easily shown that, in the limit $\gamma_1\to 0$, $5\J_2$ and $5\K_2$ are independent and $\chi^2$-distributed with 5 degrees of
freedom (d.o.f.).
In general, for the Gaussian initial conditions considered here, the joint probability density for all the variables
$\vy=(\nu,\K_{ij},\eta_k,\zeta_{lm})$ needed to define a peak and the tidal shear $\K_{ij}$ is given by a multivariate Gaussian, with
\begin{gather}
\label{eq:PDF}
Q(\vy) \equiv - \ln p(\vy) = \frac{\nu^2+\J_1^2-2\gamma_1 \nu \J_1}{2(1-\gamma_1^2)}+\frac{3}{2}\eta^2
+\frac{5}{2}\left(\frac{\K_2+\J_2+ 2 \gamma_1 \Q_2}{1-\gamma_1^2}\right) \;.
\end{gather}
Here, $\eta^2=\sum_i\eta_i^2$ is the modulus squared of the normalized density gradient. 
Note the pre-factor $+2\gamma_1$ (rather than $-2\gamma_1$) of $\Q_2$ in the last term, which follows from the definition of $\bar\zeta_{ij}(\vk)$
in terms of $\d^{(1)}(\vk)$.

\subsection{Including shear in microscopic biasing}
\label{sec:barrier}

We now wish to explicitly include the shear in the biasing prescription.
For this purpose, we first need to specify the critical density threshold for collapse, which in general depends on the tidal shear. 
In the excursion set framework, each degree of freedom adds an extra random walk.
Hence, the identification of the Lagrangian patches (proto-halos) that collapse to form halos is mapped onto the first-crossing of a
multi-dimensional barrier by multi-dimensional random walks. The first-crossing condition generically reads
\begin{equation}
\mathcal{B}(\d^{(1)},K_{2},...) = 0 \;,
\end{equation}
where $K_2 = \sigma_0 \K_2$, and we have emphasized the fact that the condition fundamentally depends on the {\it physical} fields $\d^{(1)}$,
$K_2$ and so on, rather than the {\it normalized} quantities $\nu=\frac{1}{\sigma_0}\d^{(1)}$, $\K_2=\frac{1}{\sigma_0}K_2$.
This equation can be solved for $\d^{(1)}$ and recast into the form
\begin{equation}
  \label{eq:barrier}
\d^{(1)}(\vx)=B(K_2(\vx);\;\delta_c,K_{2c},...) \;.
\end{equation}
This defines the multi-dimensional collapse barrier $B$, which generically depends on the critical density $\d_c$ for spherical collapse,
on the gravitational tidal shear $K_2(\vx)$ at the same location, a corresponding tidal shear critical threshold $K_{2c}$, and so on.
Any realistic collapse barrier $B$ asymptotes to the spherical collapse threshold $\d_c$ in the limit $\sigma_0\to 0$.
Moreover, the barrier can be measured from numerical simulations by studying the distribution of peak heights in the initial conditions.
It is convenient to label the mass scale using the ``spherical collapse significance'' $\nu_c\equiv\d_c/\sigma_0$ regardless of the form of
the barrier. This convention is ubiquitous in the literature and we shall adopt it here as well.
Note, however, that the actual height $(B/\sigma_0)$ generally differs from $\nu_c$ when $B \neq \d_c$. 
As a three-dimensional symmetric matrix, the amplitude of the local shear is completely characterized by its three rotational invariants,
$\delta$, $K_2$ and $K_3$.
Terms of the form $K^{ij}\partial_i\partial_j\d^{(1)}$ will appear if the proto-halo profile aligns with the tidal shear.

Again, the physical barrier $B$ is independent of the normalization amplitude $\sigma_8$.
The physical reasoning is that the gravitational collapse of an entire region of size $R$ is determined by the mean density
and tidal shear within that region, 
and must be independent of the variance of the density perturbations. 
In the simplest version of the barrier, we have $B=\delta_c$, which corresponds to the spherical collapse approximation.
Deviations from spherical collapse introduce a dependence on the local tidal shear \cite{Sheth:1999su,ohta/etal:2004}. 
However, in most practical implementations of ellipsoidal collapse, the physical barrier $B$ is approximated by a ``moving'' barrier with an
explicit dependence on $\sigma_8$ \cite{sheth/mo/tormen:2001,Zhang:2005ar,Moreno:2007aw,Maggiore:2009rw,Paranjape:2012ks,Sheth:2012fc,Lapi:2013bxa},
with the exception of \cite{Chiueh:2000yp,Sheth:2001dp,Sandvik:2006ue,Castorina:2016tuc} who introduced models in which $B$ depends only on
the physical density and tidal shear.
As we will see in \S\ref{sec:NGconsistency}, this point is essential to the consistency relation of the non-Gaussian bias.

Once the barrier $B$ is specified, we can write down the ``localized'' (in the sense of a point process), microscopic
number density $\npk(\vy)$ of excursion set peaks.
For the model considered here, $\vy=(\nu,\K_{ij},\eta_k,\zeta_{lm})$ and the number density of ESP peaks on the smoothing scale $R$ is given by 
\begin{align}
\npk(\vy) &= \left(\frac{\J_1}{\gamma_1\nu_c}\right) 
\frac{(6\pi)^{3/2}}{V_*} \big\lvert{\rm det}\zeta_{ij}\big\lvert \delta_D\big[\veta\big]
\Theta_H\big[\lambda_3\big] \nonumber \\
&\qquad \times \delta_D\bigg[\nu-\frac{1}{\sigma_0}B(\delta_c,K_2,K_{2c},...)\bigg] \;,
\label{eq:npk}
\end{align}
where $V_*$ is the effective volume of a Gaussian filter of radius $R_*\equiv \sqrt{3}\sigma_1/\sigma_2$, $\veta$ is the vector of components $\eta_i$
and $\lambda_3$ is the smallest eigenvalue of the matrix with components $-\zeta_{ij}$.
The step function $\Theta_H[\lambda_3]$ necessarily implies the peak condition $\J_1 > 0$ and, in addition, the first-crossing condition $-d\delta_R/dR > 0$
for the Gaussian filter we consider here.
In the interpretation of \cite{Biagetti:2013hfa} (to be contrasted with \cite{Paranjape:2012ks}, who require $-d\delta_R/dR > -dB/dR$),
this means that the random walk $\d_R^{(1)}(\vx)$ up-crosses the collapse barrier $B$ on the halo scale $R$.
As shown in \cite{Musso:2011ck}, this is a very good approximation to the first-crossing distribution down to a halo mass $M\sim M_\star$, where
$M_\star(\tau_0)$ is the characteristic mass of halos collapsing at time $\tau_0$.
The factor $\J_1/\gamma_1\nu_c$ on the right-hand side follows from the first-crossing constraint.
The presence of an additional multiplicative factor of $\nu_c^{-1}$, which appears because we follow common practice and label the halo mass with the
variable $\nu_c$, will be crucial to our discussion of universality, cf.~\S\ref{sec:universality}. 
Moreover, we have $B/\sigma_0 \propto \sigma_8^{-1}$ since the physical barrier $B$ is independent of the fluctuation amplitude.

The multiplicity function or first-crossing distribution of this microscopic point process is given by
\begin{equation}
\label{eq:fpk}
f_\text{pk}(\nu_c) = V\Big\la \npk(\vy)\Big\ra_\vy \equiv V \bnpk \;,
\end{equation}
where $V\equiv 4\pi R^3/3$ 
is the Lagrangian volume of a dark matter halo, and the brackets denote the ensemble average 
\begin{equation}
  \label{eq:av}
\Big\la \cdot \Big\ra_\vy \equiv \int {\rm d}^\dof \vy \,\Big(\cdot\Big)\Big|_{\vy}\, p(\vy)\,.
\end{equation}
Here, $\dof=6+3+6=15$ is the total number of d.o.f. and 
$p(\vy)$ [\refeq{PDF}] is the joint PDF for the variable $\vy$, which is a multivariate Normal for Gaussian initial conditions. 
The multiplicity function $f_\text{pk}$ generally depends on the fixed collapse thresholds $\delta_c$, $K_{2c}$, ..., spectral shape $\gamma_1$,
as well as the filtering scale $R$. 
For simplicity, however, we follow standard convention and parameterize the $R(M)$-dependence through $\nu_c = \d_c/\sigma_0(R)$, and do not explicitly
write the dependences on the collapse thresholds and power spectrum shape. 
Finally, the number density $\bar n_h$ of virialized halos is related to the multiplicity function through the relation
\begin{equation}
  \label{eq:nh}
\bar n_h(M) \equiv \frac{dN_h}{d\ln M} = \frac{\bar\rho_m}{M}\, f_\text{pk}(\nu_c)\, \frac{d\nu_c}{d\ln M} =\bnpk \frac{d\nu_c}{d\ln M}\;.
\end{equation}
Since $f_\text{pk}$ and $\bnpk$ are interchangeable, we shall work with $\bnpk$ in the remainder of the paper.

All our discussion straightforwardly generalizes to arbitrary combinations of variables $\vy$ and microscopic number densities $\npk(\vy)$. 
As a rule of thumb, the microscopic number density should always be a function of the normalization amplitude through the peak height $\nu(\vx)$ solely.

\subsection{Lagrangian bias functions and shear bias parameters}

Given a \emph{microscopic {description}} for the Lagrangian halo density field $\d_h^L(\vx)$ [\refeqs{dhmicro}{dhmicroK}], we can expand this
field in Fourier space in powers of the linear matter density field, in terms of general bias functions $c_n^L(\vk_1,\cdots,\vk_n)$: 
\begin{equation}
\label{eq:Fdpkeff}
\d_h^L(\vk) = \sum_{n=1}^\infty \frac{1}{n!}
\int_{\vk_1}\dots \int_{\vk_n}\,
c_n^L(\vk_1,\dots,\vk_n)\, \Big[\delta^{(1)}(\vk_1)\dots\delta^{(1)}(\vk_n)+\cdots\Big](2\pi)^3 \delta_D\!\big(\vk_{1\dots n}-\vk\big) \;,
\end{equation}
where $\vk$ is the Fourier wavemode, $\vk_{1\dots n}=\vk_1+\dots +\vk_n$, and the ellipsis in the square bracket stands for terms containing
$n-2$, $n-4, \dots$ powers of the density field. 
These lower-order terms are present because each integrand corresponds to a sum of $n$-th order renormalized operators written in Fourier space,
which contain lower-order operators as counter terms (multiplied by powers of moments $\sigma_i^2,\,\sigma_i^2 \sigma_j^2, \dots$).  

The Lagrangian bias functions $c_n^L$ follow from the equality \cite{Matsubara:2011ck}
\begin{equation}
\label{eq:cnL}
c_n^L(\vk_1,\dots,\vk_n) \equiv \frac{1}{\bnpk}\Big\langle\npk(\vy) \mathcal{D}_\vy(\vk_1)\dots \mathcal{D}_\vy(\vk_n)\Big\rangle_\vy \;.
\end{equation}
Our choice of ordering signifies that the differential operator $\mathcal{D}_\vy(\vk)$ acts on the density PDF $p(\vy)$ and not on $\npk(\vy)$
(see Eq.~(\ref{eq:av})).
The functional form of $\mathcal{D}_\vy(\vk)$ depends on the details of the point process \cite{Matsubara:2011ck}.
For the variables considered here (peak constraint and tidal shear), it is 
\begin{equation}
\mathcal{D}_\vy(\vk) = W_R(k) \left(\frac{i}{\sigma_1} k_i \frac{\partial}{\partial\eta_i}
+\frac{1}{\sigma_0} \frac{k_i k_j}{k^2}\frac{\partial}{\partial\K_{ij}}-\frac{1}{\sigma_2} k_i k_j \frac{\partial}{\partial\zeta_{ij}} 
\right) \;,
\end{equation}
where repeated indices are summed. 
Although $\K_{ij}$ and $\zeta_{ij}$ only have 6 independent components, it is convenient to keep the redundancy for computational purposes, so
that index summations run from 1 to 3 unrestrictedly.
Moreover, we will generically denote the low-pass filtering kernel as $W_R(k)$, but the reader should bear in mind that we have implicitly assumed
a Gaussian filter throughout this paper.

As discussed in \cite{Desjacques:2012eb,Lazeyras:2015giz,Matsubara:2016wth}, this expansion is equivalent to a series expansion of the 
microscopic peak number density $\dpk(\vx)\equiv \frac{\npk(\vx)}{\bnpk}-1$ in orthogonal polynomials. As a result, the peak bias parameters are
all renormalized. 
Since, in peak theory (and, more generically, in any 
``microscopic'' Lagrangian bias description), 
the non-Gaussian bias amplitude is an integral of the second-order Lagrangian bias function $c_2^L$ \cite{Matsubara:2011ck,Desjacques:2013qx}
(see also Appendix \ref{app:ngbias}),
\begin{equation}
b_\phi \equiv \int_{\vq}\! c_2^L(\vq,-\vq)\,\plin(q) \;,
\label{eq:bphi}
\end{equation}
we need to evaluate second derivatives of $p(\vy)$ w.r.t. $\eta_i$, $\K_{ij}$ and $\zeta_{ij}$ along the lines of \cite{Matsubara:2016wth}. 
The technical details can be found in Appendix \ref{app:details}. 
After some manipulations, we arrive at
\begin{align}
\mathcal{D}&(\vk_1)\mathcal{D}(\vk_2)\, p(\vy) = 
W_R(k_1)W_R(k_2)\bigg\{-\frac{2}{\sigma_1^2}\big(\vk_1\cdot\vk_2)
\bigg(1+\frac{2}{3}\eta^2\frac{\partial}{\partial(\eta^2)}\bigg)\frac{\partial}{\partial(\eta^2)} \nonumber \\
&\quad 
+\frac{1}{\sigma_0^2}\bigg[\frac{\partial^2}{\partial\nu^2}+\Big(3\big(\kvh_1\cdot\kvh_2\big)^2-1\Big) 
\bigg[\frac{\partial}{\partial \K_2} + \frac{2}{5}\bigg(\K_2 \frac{\partial^2}{\partial \K_2^2}
+ \Q_2\frac{\partial^2}{\partial \K_2\partial \Q_2}+\frac{1}{4}\J_2\frac{\partial^2}{\partial \Q_2^2}\bigg)\bigg] \nonumber \\
&\quad + \frac{1}{\sigma_2^2} k_1^2 k_2^2\bigg[ \frac{\partial^2}{\partial \J_1^2}
+\Big(3\big(\kvh_1\cdot\kvh_2\big)^2-1\Big) \bigg[\frac{\partial}{\partial \J_2} + \frac{2}{5}\bigg(\J_2 \frac{\partial^2}{\partial \J_2^2}
+\Q_2\frac{\partial^2}{\partial \J_2\partial \Q_2}+\frac{1}{4}\K_2\frac{\partial^2}{\partial \Q_2^2}\bigg)\bigg] \nonumber \\
&\quad +\frac{1}{\sigma_0\sigma_2}\Big(k_1^2+k_2^2\Big)\bigg[\frac{\partial^2}{\partial\nu\partial \J_1}-
\Big(3\big(\kvh_1\cdot\kvh_2\big)^2-1\Big)\bigg[ \frac{1}{2}\frac{\partial}{\partial \Q_2}
+\frac{2}{5} \bigg(\Q_2 \frac{\partial^2}{\partial \J_2\partial \K_2} \nonumber \\
& \quad +\frac{1}{2}\J_2 \frac{\partial^2}{\partial \J_2\partial \Q_2} 
+ \frac{1}{2}\K_2 \frac{\partial^2}{\partial \K_2\partial \Q_2} +\frac{1}{4} \Q_2 \frac{\partial^2}{\partial \Q_2^2}\bigg)\bigg]\bigg\} 
p(\vy)\;.
\label{eq:DDav}
\end{align}
Here, we have not written the terms that vanish under the integral ${\rm d}^D \vy$. 
Multiplying the result by the ESP number density Eq.~(\ref{eq:npk}), integrating over the variables $\vy$ and dividing by $\bnpk$ as in Eq.~(\ref{eq:cnL}),
we obtain
\begin{align}
  \label{eq:c2Lshear}
c_2^L(\vk_1,\vk_2)  &= \biggl\{b_{20}^L + b_{11}^L \left(k_1^2+k_2^2\right) 
+ b_{02}^L k_1^2 k_2^2-2 \chi_1^L \left(\vk_1\cdot\vk_2\right) \\ 
& \qquad  + \Bigl[b_{\K_2}^L+b_{\mathcal{Q}_2}^L\left(k_1^2+k_2^2\right)+b_{\J_2}^Lk_1^2k_2^2\Bigr]
\Bigl[3\big(\kvh_1\cdot\kvh_2\big)^2 -1\Bigr]\biggr\}\, W_R(k_1) W_R(k_2) \nonumber \;.
\end{align}
Here, the renormalized Lagrangian peak biases $b_{ij}^L$ and $\chi_1^L$ are given by (e.g., \cite{Lazeyras:2015giz})
\begin{align}
b_{ij}^L &= \frac{1}{\sigma_0^i\sigma_2^j\bnpk}\int\!\!{\rm d}^{\dof}\vy\,\npk(\vy) H_{ij}(\nu,\J_1)\, p(\vy) \label{eq:bij} \\
\chi_k^L &= \frac{1}{\sigma_1^{2k}\bnpk}\int\!\!{\rm d}^{\dof}\vy\,\npk(\vy) L_k^{(1/2)}\!(3\eta^2)\, p(\vy) \label{eq:chik}\;,
\end{align}
where $H_{ij}(x,y)$ and $L_n^{(\alpha)}(x)$ are bivariate Hermite and generalized Laguerre polynomials, respectively.
Note that $H_{i0}(\nu,\J_1)\equiv H_i(\nu)$ and $H_{0j}(\nu,\J_1)\equiv H_j(\J_1)$.
These polynomials appear since $\nu, J_1$ are distributed as a bivariate Normal distribution, while $3\eta^2$ follows a $\chi^2$ distribution with 3 d.o.f..

The inclusion of shear leads to two additional terms in $c_2^L$ (see the second line of Eq.~(\ref{eq:c2Lshear})), which are controlled by the Lagrangian
bias parameters $b_{\K_2}^L$ and $b_{\Q_2}^L$ and are not present in previously considered peak bias expansions \cite{Desjacques:2012eb,Matsubara:2016wth}.
The former, for which there are predictions from \cite{Sheth:2012fc,castorina/etal:2017} based on excursion sets, has been recently found to be non-vanishing
\cite{Modi:2016dah}.
The latter is due to the correlation between $\K_{ij}$ and $\zeta_{ij}$ and has only been considered in \cite{castorina/etal:2017} (it is also included in the
list of second-order higher-derivative bias contributions in \S~2.6 of \cite{Desjacques:2016bnm}). 
These bias parameters are given by the following integrals over the localized peak density
($b_{\K_2}^L\propto c_2$ in the notation of \cite{Sheth:2012fc}, and $b_{\Q_2}^L\propto c_{21}$ in the notation of \cite{castorina/etal:2017})
\begin{align}
  b_{\K_2}^L &= \frac{1}{\sigma_0^2\bnpk}\int\!\!{\rm d}^{\dof}\vy\,\npk(\vy)\,
  \frac{5}{2}\left[\frac{\K_2+2\gamma_1 \Q_2+\gamma_1^2 \J_2}{(1-\gamma_1^2)^2}-\frac{1}{(1-\gamma_1^2)}\right]\, p(\vy) \\
  b_{\Q_2}^L &= \frac{1}{\sigma_0\sigma_2\bnpk}\int\!\!{\rm d}^{\dof}\vy\,\npk(\vy)\,
  \left(-\frac{5}{2}\right)\left[\frac{(1+\gamma_1^2)\Q_2+\gamma_1(\K_2+\J_2)}{(1-\gamma_1^2)^2}-\frac{\gamma_1}{(1-\gamma_1^2)}\right]\, p(\vy) \;.
\end{align}
whereas, for sake of completeness, ($b_{\J_2}^L\equiv\omega_{10}^L$ in the notation of \cite{Desjacques:2016bnm}):
\begin{equation}
  b_{\J_2}^L = \frac{1}{\sigma_2^2\bnpk}\int\!\!{\rm d}^{\dof}\vy\,\npk(\vy)\,
  \frac{5}{2}\left[\frac{\J_2+2\gamma_1 \Q_2+\gamma_1^2 \K_2}{(1-\gamma_1^2)^2}-\frac{1}{(1-\gamma_1^2)}\right]\, p(\vy) \;.
\end{equation}
In the limit $\gamma_1\to 0$, the integrand of $b_{\K_2}^L$ and $b_{\J_2}^L$ converges to respectively $-L_1^{(3/2)}(\frac{5}{2}\K_2)$ and
$-L_1^{(3/2)}(\frac{5}{2}\J_2)$ as expected, while the integrand of $b_{\Q_2}^L$ simplifies to $-\frac{5}{2} \Q_2$.
Therefore, $b_{\Q_2}^L$ is generally non-zero in the limit where $\bar{\K}_{ij}$ and $\bar{\zeta}_{ij}$ are decoupled even though, for
$\gamma_1=0$, the probability density $p(\vy)$ does not explicitly depend on $\Q_2$. 
In the particular case of spherical collapse, there is no dependence on shear ($B=\delta_c$) so that both $b_{\Q_2}^L$ and $b_{\K_2}^L$ vanish.

\section{Bias consistency relations and universality}
\label{sec:bias}

We will now demonstrate that the consistency relation for the non-Gaussian 
bias 
\begin{equation}
\label{eq:PBS2}
b_\phi = \frac{\partial\ln\bar n_h}{\partial\ln\sigma_8} \;,
\end{equation}
holds for microscopic Lagrangian bias desciptions with generalized collapse barriers as long as the dependence on variables other than density
(such as tidal shear) is explicitly modeled. 
We will also elucidate under which circumstances the non-Gaussian bias consistency relation is valid, and discuss the origin of 
(non-)universality in halo mass functions.

\subsection{Consistency relation for the non-Gaussian bias}
\label{sec:NGconsistency}

We first demonstrate that \refeq{PBS2} holds for any generic moving and stochastic barrier $B(\delta_c,K_2,K_{2c},...)$ which incorporates
explicitly the effect of the tidal shear. 
Firstly, we calculate $b_\phi$ given by \refeq{bphi} with aid of the following relations:
\begin{align}
\int_\vq \bigg\{p(\vy)^{-1} \Big(\mathcal{D}(\vq)\mathcal{D}(-\vq)\Big)_{[\nu,\J_1]}\, p(\vy)\bigg\} \plin(q) 
&= \frac{\nu^2+\J_1^2-2\gamma_1\nu \J_1}{(1-\gamma_1^2)}-2 \\
\int_\vq \bigg\{p(\vy)^{-1} \Big(\mathcal{D}(\vq)\mathcal{D}(-\vq)\Big)_{[\eta^2]}\, p(\vy)\bigg\} \plin(q) 
&= 3\eta^2 -3 \;,
\end{align}
where the notation $(...)_{[X]}$ signifies that the calculation is restricted to the variables $X$. 
For $\K_2$, $\mathcal{Q}_2$ and $\J_2$, we find:
\begin{equation}
\int_\vq \bigg\{p(\vy)^{-1} \Big(\mathcal{D}(\vq)\mathcal{D}(-\vq)\Big)_{[\K_2,\mathcal{Q}_2,\J_2]}\, p(\vy)\bigg\} \plin(q) 
=\frac{5(\K_2+\J_2+2\gamma_1 \mathcal{Q}_2)}{(1-\gamma_1^2)}-10
\end{equation}
Collecting these various results, we arrive at
\begin{align}
b_\phi & = \int_\vq\! c_2^L(\vq,-\vq)\, \plin(q) \nonumber \\
&=\frac{1}{\bnpk}\int_\vq \Big\langle\npk\mathcal{D}(\vq)\mathcal{D}(-\vq)\Big\rangle \plin(q)
\nonumber \\
&= \frac{1}{\bnpk}\int_\vq \int\!{\rm d}^{\dof}\vy\,\npk(\vy)\mathcal{D}(\vq)\mathcal{D}(-\vq) p(\vy)\,\plin(q) 
\nonumber \\
&= \frac{1}{\bnpk}\int\!{\rm d}^{\dof}\vy\,\npk(\vy)p(\vy) \left\{\int_\vq\bigg[p(\vy)^{-1}\mathcal{D}(\vq)\mathcal{D}(-\vq) p(\vy)\bigg]\plin(q)\right\} 
\nonumber \\
&= \frac{1}{\bnpk}\int\!{\rm d}^{\dof}\vy\,\npk(\vy) p(\vy)\bigg(2 \tilde Q(\vy)-\dof\bigg) \label{eq:LHS} \;.
\end{align}
where, again, $\dof= 15$ is the total number of degrees of freedom, and $\tilde Q(\vy)$ is the quadratic form that appears in
the probability density $p(\vy)$ (see Eq.~(\ref{eq:PDF})).

Secondly, we massage the logarithmic derivative of $\bar n_h$ w.r.t. $\sigma_8$, which reads
\begin{equation}
\frac{\partial}{\partial\ln\sigma_8}\bigg(\bnpk(\nu_c)\frac{d\nu_c}{d\ln M}\bigg) 
\equiv \frac{\partial}{\partial\ln\sigma_8}\left\{\int\!{\rm d}^{\dof}\vy\,\npk(\vy) p(\vy)\,\frac{d\nu_c}{d\ln M}\right\} \;.
\end{equation}
To proceed further, we exploit the fact that the Dirac distribution satisfies $\d_D(a x) = |a|^{-1}\delta_D(x)$ and write
$\delta_D[\nu(\vx)-B/\sigma_0]d\nu=\delta_D[\delta(\vx)-B]d\delta$. On performing the integral over $\delta$, we are left with an integral
over the remaining $\dof-1$ variables $\vw$ whose integrand is uniformly continuous on the relevant $\sigma_8\times \vw$
domain. Therefore, we can at this point exchange the derivative w.r.t. $\sigma_8$ and the integral.
Firstly, the measure $d^{\dof-1}\vw$ is proportional to $\sigma_8^{-\dof+1}$, so that its logarithmic derivative w.r.t. $\sigma_8$ returns
$-\dof+1$. Secondly, $p(\vy)=p(\vw,\frac{B}{\sigma_0})$ depends on $\sigma_8$ through $\tilde Q(\vw,\frac{B}{\sigma_0})\propto \sigma_8^{-2}$.
Therefore, the logarithmic derivative of $p(\vy)$ gives $2\tilde Q(\vw,\frac{B}{\sigma_0})$.
Finally, the microscopic peak number density $\npk(\vy)$ does not depend on $\sigma_8$ once the Dirac distribution has been integrated out.
However, the Jacobian $\frac{d\nu_c}{dM}\propto \sigma_8^{-1}$ yields an additive contribution of $-1$ to the logarithmic derivative.
Summing up all the contributions, we obtain
\begin{equation}
\label{eq:RHS}
\frac{\partial}{\partial\ln\sigma_8}\bigg(\bnpk(\nu_c)\frac{d\nu_c}{d\ln M}\bigg)  = 
\int\!{\rm d}^{\dof}\vy\,\npk(\vy) p(\vy)\bigg(2\tilde Q(\vy)-\dof\bigg)\frac{d\nu_c}{d\ln M}
\end{equation}
where we have made $\npk(\vy)$ etc. explicit again.
Therefore, combining Eqs. (\ref{eq:LHS}) and (\ref{eq:RHS}), we arrive at
\begin{align}
\frac{\partial\ln\bar n_h}{\partial\ln\sigma_8} - b_\phi
&= \frac{1}{\bnpk}\frac{\partial}{\partial\ln\sigma_8}\bigg(\bnpk(\nu_c)\frac{d\nu_c}{d\ln M}\bigg) 
\left(\frac{d\nu_c}{d\ln M}\right)^{-1} \nonumber \\ 
& \qquad - \frac{1}{\bnpk}\int\!{\rm d}^{\dof}\vy\,\npk(\vy) p(\vy)\bigg(2 \tilde Q(\vy)-\dof\bigg) 
\nonumber \\
&= 0 \;,
\end{align}
which proves the peak-background split consistency relation Eq.~(\ref{eq:PBS}).

This result easily generalizes to more sophisticated Lagrangian bias prescriptions. In particular, adopting a filter function $W_R(k)$
different from Gaussian leads to the introduction of an additional variable $\mu\propto -d\delta_R/dR$. 
The localized number density becomes
\begin{align}
  \npk(\vy) &= \left(\frac{\mu}{\gamma_{\nu\mu}\nu_c}\right) 
\frac{(6\pi)^{3/2}}{V_*} \big\lvert{\rm det}\zeta_{ij}\big\lvert \delta_D\big[\veta\big]
\Theta_H\big[\lambda_3\big] \Theta_H\big[\mu\big] \nonumber \\
&\qquad \times \delta_D\bigg[\nu-\frac{1}{\sigma_0}B(\delta_c,K_2,K_{2c},...)\bigg] \;,
\label{eq:npkmu}
\end{align}
where $\mu$, $\zeta_{ij}$, $\veta$, $\lambda_3$, $\nu$ and $K_2$ are all functions of $\vx$.
Here, $\gamma_{\nu\mu}$ is the correlation coefficient between $\nu$ and $\mu$.
There is now one additional d.o.f., that is,  $\dof=16$, but the extra multiplicative factor of $\mu/\nu_c$ does not change under a rescaling
of $\sigma_8$. Therefore, Eq.~(\ref{eq:PBS}) also holds in this case. 

However, the consistency relation is violated as soon as the collapse barrier $B$ explicitly depends on $\sigma_8$, as is the case in many
approximations in the literature. A straightforward calculation indeed shows that
\begin{equation}
\frac{\partial\ln\bar n_h}{\partial\ln\sigma_8} - b_\phi
= -\frac{1}{\sigma_0\bnpk}\int\!{\rm d}^{\dof}\vy\,\npk(\vy)\left(\frac{\partial B}{\partial\ln\sigma_8}\right) H_1(\nu)\,
p(\vy)\;.
\end{equation}
for a general barrier. In the particular case of the square-root barrier
\begin{equation}
\label{eq:Bsqrt}
B = \delta_c + \sigma_0 \beta \;,
\end{equation}
used in the original implementation of the ESP approach \cite{Paranjape:2012jt}, the logarithmic derivative of $B$ relative to $\sigma_8$
returns the stochastic field $\beta$ that describes the scatter around the mean barrier, which is modeled as an additional variable with
non-vanishing 1-point covariance solely. In this case, $\frac{\partial\ln\bar n_h}{\partial\ln\sigma_8} - b_\phi\ne 0$ especially for
halos with mass $M\lesssim M_\star$ \cite{Biagetti:2015exa}.
This is clearly at odds with numerical evaluations of $b_\phi$ and $\frac{\partial\ln\bar n_h}{\partial\ln\sigma_8}$ for low mass halos,
which agree very well regardless of the universality of the halo mass function \cite{biagetti/lazeyras/etal:2016}.

\subsection{Universality of the halo mass function}
\label{sec:universality}

We now turn to the universality of the halo mass function, for which the first-crossing constraint is critical.
As outlined in the introduction, universality is usually assumed to refer to a mass function of the form Eq.~(\ref{eq:nh}) for which the
multiplicity function $f_\text{pk}$ is a function of $\nu_c$ solely.
However, since for peaks $f_\text{pk}$ also depends on ratios of spectral moments $\sigma_i$, for example, we shall consider \refeq{universal}
as the definition of universality instead:
\begin{equation}
  \frac{\partial\ln\bar n_h}{\partial\ln\sigma_8} = b_1^L \d_c \;.
\label{eq:universal_this}
\end{equation}
This is the relation tested in, e.g., \cite{tinker/robertson/etal:2010,biagetti/lazeyras/etal:2016} with numerical simulations.
We emphasize, however, that such a relation does not arise from a 
peak-background split which provides the consistency relation we showed in Sec.~\ref{sec:NGconsistency} and as such, does not have
to hold for all tracers.

Our starting point is the halo mass function Eq.~(\ref{eq:nh}).
Since the average number density $\bnpk(\nu_c)$ is a function of $\sigma_8$ through $\nu_c\propto \sigma_8^{-1}$ solely, 
taking the logarithmic derivative of $\bar n_h$ yields
\begin{align}
\frac{\partial\ln\bar n_h}{\partial\ln\sigma_8} &= \frac{\partial\ln\bnpk}{\partial\ln\sigma_8} -1 \nonumber \\
& = - \frac{\partial\ln\bnpk}{\partial\ln\nu_c} -1 \nonumber \\
&= \bigg\{1 - \big(\gamma_1\bnpk\big)^{-1}
\frac{\partial}{\partial\nu_c}\int\!{\rm d}^{\dof}\vy\, \npk(\vy) p(\vy) \bigg\} -1 \nonumber \\
&= -\big(\gamma_1\bnpk\big)^{-1}
\int\!{\rm d}^{\dof}\vy\, \left(\frac{\partial}{\partial\nu_c}\npk(\vy)\right) p(\vy) \;.
\end{align}
In the third equality, the first-crossing constraint brings an additional factor of $+1$ (through the multiplicative factor
$(\J_1/\gamma_1\nu_c)$) which exactly cancels the factor of $-1$ returned by the Jacobian $\frac{d\nu_c}{dM}$.
Since the microscopic number density $\npk(\vy)$ depends on $\nu_c$ through the Dirac factor $\delta_D(\nu-\frac{B}{\sigma_0})$,
we must consider
\begin{align}
\frac{\partial}{\partial\nu_c}\npk(\vy) &\propto \frac{\partial}{\partial\nu_c}\delta_D(\nu-B/\sigma_0) \nonumber \\
&= \sigma_0^2\frac{\partial}{\partial\delta_c}\delta_D(\delta-B) \nonumber \\
&= -\sigma_0^2\frac{\partial}{\partial\delta}\delta_D(\delta-B)\frac{\partial B}{\partial\delta_c} \nonumber \\
&= -\frac{\partial}{\partial\nu}\delta_D(\nu-B/\sigma_0)\frac{\partial B}{\partial\delta_c} \;.
\end{align}
Substituting this result in the above expression and integrating by parts, the logarithmic derivative of $\bar n_h$ becomes
\begin{align}
  \frac{\partial\ln\bar n_h}{\partial\ln\sigma_8} &= -\frac{\nu_c}{\bnpk}
  \int\!{\rm d}^{\dof}\vy\,\npk(\vy) 
\left(\frac{\partial B}{\partial\delta_c}\right)
\left(\frac{\partial}{\partial\nu}p(\vy)\right) \;.
\end{align}
For peaks, the linear Lagrangian LIMD bias is $b_1^L\equiv b_{10}^L$, with
\begin{equation}
  \label{eq:LIMDb1}
b_{10}^L = \frac{1}{\sigma_0\bnpk}\int\!\!{\rm d}^{\dof}\vy\,\npk(\vy) H_{10}(\nu,\J_1)\, p(\vy) \;.
\end{equation}
Together with Rodrigues' formula for Hermite polynomials, we eventually arrive at
\begin{equation}
\frac{\partial\ln\bar n_h}{\partial\ln\sigma_8} - \delta_c b_{10}^L 
= \frac{\nu_c}{\bnpk}\int\!{\rm d}^{\dof}\vy\,\npk(\vy)\left(\frac{\partial B}{\partial\delta_c}-1\right) H_{10}(\nu,\J_1)p(\vy)\;,
\end{equation}
which vanishes only if a shift $\Delta\d_c$ in the spherical collapse threshold changes the collapse barrier by precisely $\Delta B=\Delta\d_c$.
Therefore, the barrier is necessarily linear in $\d_c$, i.e. 
\begin{equation}
  B=\delta_c + g(K_2,K_{2c},...)\;.
\end{equation}
Since, {\it by definition}, a long-wavelength density perturbation $\d_\ell^{(1)}$ translates the collapse barrier by an amount
$\Delta B = - \d_\ell^{(1)}$, universality implies
\begin{equation}
  \label{eq:universalcondition}
  \Delta\d_c = - \d_\ell^{(1)} \;.
\end{equation}
Such a condition would be violated if the barrier is of the form $B=\sqrt{\d_c^2 + K_2/K_{2c}}$ for instance.
By contrast, it would be satisfied for a barrier $B=\sqrt{\d_c^2 + K_3 / K_{3c}}$ if $K_3\propto{\rm tr}(K_{ij}^3)$ is replaced by its ensemble
average $\big\la K_3\big\ra=0$.
This illustrates the danger of using effective ``moving'' barriers in which some of the physical variables are replaced by their ensemble
average ($\sigma_8$-dependent) values.

When universality holds, derivatives of the halo mass function w.r.t. the long-wavelength perturbation are equivalent to
derivatives w.r.t. the spherical collapse threshold $\d_c$.
At the level of the function $F(\d^{(1)}(\vx),K_{ij}(\vx),...)$ defined in \refeq{dhmicroK}, the requirement $\Delta\d_c = - \d_\ell^{(1)}$
immediately leads to \refeq{dfdc} as advertised in the introduction. \refeq{dfdc} generalizes to 
\begin{equation}
  \label{eq:dfdcn}
  \frac{\partial^n F}{\partial\big[\d_\ell^{(1)}\big]^n} = (-1)^n \frac{\partial^n F}{\partial\d_c^n} \;,
\end{equation}
so that all the Lagrangian LIMD bias parameters can be computed from the derivatives of the function $F$ w.r.t. $\d_c$ rather than
$\d_\ell^{(1)}$.
Note that there is no restriction on the functional form of the multiplicity function, which can be a function of $\nu_c$ alone or involve also 
ratios of the spectral moments $\sigma_i$, as is the case for peaks for instance.

The analysis of \cite{despali/giocoli/etal:2015,biagetti/lazeyras/etal:2016} shows that the universality of the halo mass function strongly
depends on the halo definition adopted. In particular, while FoF (friends-of-friends) halos are very close to universal as they were found to
satisfy \refeq{universal_this} for all resolved halo masses, 
SO halos break universality significantly at mass $M\lesssim M_\star$.
Therefore, the barrier not only depends on the physical fields, but also on the characteristics of the halo finder. 
However, the barrier will always converge towards the spherical collapse threshold $\delta_c$ in the limit of large halo mass. Therefore,
\refeq{universal_this} always holds in the high peak limit regardless of the halo finder.

\section{Conclusion}
\label{sec:conclusions}

In the framework of microscopic Lagrangian halo descriptions, we have demonstrated under which circumstances the consistency relation for the
non-Gaussian bias and the universality of the halo mass function hold.
The consistency relation \refeq{PBS} is valid as long as the barrier $B$ depends only on the physical fields and explicitly includes the effect
of the tidal shear --- or, more generally, the physical sources for the scatter --- rather than a parameterized scatter.
Similarly, the halo mass function is universal provided that a long-wavelength, linear density perturbation $\d_\ell^{(1)}$ changes
the spherical collapse threshold by $\Delta \d_c=  - \d_\ell^{(1)}$
or, equivalently, the response of the function $F$ defining the biased tracers to $\d_\ell^{(1)}$ is given by \refeq{dfdc}.
Our results apply to any Lagrangian bias prescription.

Effective ``moving'' barriers do not generally satisfy the consistency relation.
In the particular case of the square-root barrier $B = \delta_c + \sigma_0 \beta$, the scatter is described by a stochastic field $\beta$ assumed
uncorrelated with the other variables.
While it leads to a universal mass function, it violates the consistency relation because $B$ explicitly depends on $\sigma_8$
(unless $\beta\propto \sigma_0^{-1}$, which can be trivially absorbed in the value of $\d_c$). 
``Microscopic'' barriers including the tidal shear bring two additional, second-order Lagrangian bias parameters which ensure that the consistency
relation is satisfied. We provide analytic expressions for these tidal shear Lagrangian bias parameters.

While the non-Gaussian bias consistency relation is valid regardless of the halo finding algorithm, the universality of the halo mass function
strongly depends on the halo identification. The findings of \cite{biagetti/lazeyras/etal:2016} suggest that, whereas FoF halos are described by
a barrier linear in $\delta_c$, this is not the case of SO halos. It would be interesting to understand better how the functional form of the
barrier is affected by the halo identification.

\acknowledgments

V.D. acknowledges support by the Israel Science Foundation (grant no. 1395/16).
D.J. acknowledges support from National Science Foundation grant (AST-1517363).
F.S. is supported by the Marie Curie Career Integration Grant  (FP7-PEOPLE-2013-CIG) ``FundPhysicsAndLSS,'' and Starting Grant (ERC-2015-STG 678652)
``GrInflaGal'' from the European Research Council.

\appendix

\section{Further details on the derivation of the shear bias}
\label{app:details}

Beginning with $\eta_i$, we have
\begin{align}
\frac{\partial }{\partial\eta_i} p(\vy) &= 2 \eta_i \frac{\partial}{\partial(\eta^2)} p(\vy) \nonumber \\
\frac{\partial^2}{\partial\eta_i\partial\eta_j} p(\vy) &= \left( 2 \delta_{ij} \frac{\partial}{\partial(\eta^2)} + 4 \eta_i \eta_j 
\frac{\partial^2}{\partial(\eta^2)^2}\right) p(\vy) \;.
\end{align}
where $\eta^2 \equiv \sum_i \eta_i^2$. For the jointly distributed variables $\K_{ij}$ and $\zeta_{ij}$, we use relations similar
to those derived by \cite{Matsubara:2016wth}, 
\begin{equation}
\frac{\partial}{\partial\zeta_{ij}}p(\vy) = \left(\frac{\partial \J_1}{\partial\zeta_{ij}}\frac{\partial}{\partial \J_1} +
\frac{\partial \J_2}{\partial\zeta_{ij}}\frac{\partial}{\partial \J_2} + \frac{\partial \Q_2}{\partial\zeta_{ij}}\frac{\partial}{\partial \Q_2}\right)
p(\vy) \;,
\end{equation}
where
\begin{equation}
\frac{\partial \J_1}{\partial\zeta_{ij}} = -\delta_{ij} \;, \qquad
\frac{\partial \J_2}{\partial\zeta_{ij}} = 3 \bar\zeta_{ij} \;, \qquad
\frac{\partial \Q_2}{\partial\zeta_{ij}} = \frac{3}{2} \bar \K_{ij} \;.
\end{equation}
Furthermore, the following relation turns out to be useful:
\begin{equation}
\frac{\partial\bar\zeta_{ij}}{\partial\zeta_{kl}} = \delta_{ik}\delta_{jl} - \frac{1}{3}\delta_{ij}\delta_{kl} \;.
\end{equation}
Armed with these results, we obtain at first order
\begin{align}
\frac{\partial}{\partial\zeta_{ij}}p(\vy) &= \left(-\delta_{ij} \frac{\partial}{\partial \J_1} + 3\bar\zeta_{ij}\frac{\partial}{\partial \J_2}
+\frac{3}{2}\bar \K_{ij}\frac{\partial}{\partial \Q_2}\right) p(\vy) \nonumber \\
\frac{\partial}{\partial \K_{ij}}p(\vy) &= \left(\delta_{ij} \frac{\partial}{\partial \nu} + 3\bar \K_{ij}\frac{\partial}{\partial \K_2}
+\frac{3}{2}\bar \zeta_{ij}\frac{\partial}{\partial \Q_2}\right) p(\vy) \;.
\end{align}
At second order, we find
\begin{align}
\frac{\partial^2}{\partial\zeta_{ij}\partial\zeta_{kl}}p(\vy) &=
\bigg[\delta_{ij}\delta_{kl}\frac{\partial^2}{\partial \J_1^2}+ \Big(3\delta_{ik}\delta_{jl}-\delta_{ij}\delta_{kl}\Big) \frac{\partial}{\partial \J_2}
-3 \Big(\delta_{ij}\bar\zeta_{kl}+\bar\zeta_{ij}\delta_{kl}\Big)\frac{\partial^2}{\partial \J_1\partial \J_2} \nonumber \\
&\quad -\frac{3}{2} \Big(\delta_{ij}\bar \K_{kl}+\bar \K_{ij}\delta_{kl}\Big)\frac{\partial^2}{\partial \J_1\partial \Q_2}
+ 9\bar\zeta_{ij}\bar\zeta_{kl}\frac{\partial^2}{\partial \J_2^2} +\frac{9}{2}\Big(\bar\zeta_{ij}\bar \K_{kl}+\bar \K_{ij}\bar \zeta_{kl}\Big)
\frac{\partial^2}{\partial \J_2 \partial \Q_2} \nonumber  \\ 
&\quad + \frac{9}{4} \bar \K_{ij}\bar \K_{kl} \frac{\partial^2}{\partial \Q_2^2}\bigg] p(\vy) \;.
\end{align}
The second derivative w.r.t. $\K_{ij}$ and $\K_{kl}$ is identical, except for $\bar\zeta_{ij}$ being replaced by $\bar \K_{ij}$, $\J_1$ by $-\nu$ and
$\J_2$ by $\K_2$. Moreover, the cross-derivative reads
\begin{align}
\frac{\partial^2}{\partial\zeta_{ij}\partial \K_{kl}}p(\vy) &= \bigg[-\delta_{ij}\delta_{kl} \frac{\partial^2}{\partial\nu\partial \J_1}
-3\delta_{ij}\bar \K_{kl} \frac{\partial^2}{\partial \K_2\partial \J_1}-\frac{3}{2}\delta_{ij}\bar \zeta_{kl}\frac{\partial^2}{\partial \Q_2\partial \J_1}
+3 \bar\zeta_{ij}\delta_{kl} \frac{\partial^2}{\partial\nu\partial \J_2} \nonumber \\
&\quad + \frac{3}{2} \bar \K_{ij}\delta_{kl} \frac{\partial^2}{\partial\nu\partial \Q_2}
+ \frac{1}{2}\Big(3\delta_{ik}\delta_{jl}-\delta_{ij}\delta_{kl}\Big) \frac{\partial}{\partial \Q_2} 
+9\bar\zeta_{ij}\bar \K_{kl} \frac{\partial^2}{\partial \J_2 \partial \K_2} \nonumber \\ 
&\quad + \frac{9}{2} \bar\zeta_{ij}\bar\zeta_{kl}\frac{\partial^2}{\partial \J_2\partial \Q_2}
+\frac{9}{2} \bar \K_{ij} \bar \K_{kl} \frac{\partial^2}{\partial \K_2\partial \Q_2}
+ \frac{9}{4} \bar \K_{ij}\bar\zeta_{kl} \frac{\partial^2}{\partial \Q_2^2}\bigg] p(\vy) \;.
\end{align}
There are no cross-terms between $\eta_i$ and $\zeta_{ij}$ or $\K_{ij}$ at second order (they arise at third order). 
Since the density field is assumed to be isotropic, the Lagrangian bias functions Eq.~(\ref{eq:cnL}) are invariant under rotations of the coordinates. 
Therefore, we follow \cite{Matsubara:2016wth} and replace products of variables by their angular averages. In particular, 
\begin{gather}
\big\langle \eta_i\eta_j\big\rangle_\Omega = \frac{1}{3}\delta_{ij} \eta^2 \;,\qquad
\big\langle \bar\zeta_{ij}\bar\zeta_{kl}\big\rangle_\Omega = \frac{1}{15} \left(\delta_{ik}\delta_{jl}+\delta_{il}\delta_{jk}
-\frac{2}{3}\delta_{ij}\delta_{kl}\right)\J_2 \;, \nonumber \\
\big\langle \bar\zeta_{ij}\bar \K_{kl}\big\rangle_\Omega = \frac{1}{15} \left(\delta_{ik}\delta_{jl}+\delta_{il}\delta_{jk}
-\frac{2}{3}\delta_{ij}\delta_{kl}\right) \Q_2 \;,
\end{gather}
in addition to $\la\bar\zeta_{ij}\ra_\Omega=\la\bar \K_{ij}\ra_\Omega=0$. We find
\begin{align}
\left\langle\frac{\partial^2}{\partial\zeta_{ij}\partial\zeta_{kl}}\right\rangle_\Omega p(\vy) &=
\bigg[\delta_{ij}\delta_{kl}\frac{\partial^2}{\partial \J_1^2}+ \Big(3\delta_{ik}\delta_{jl}-\delta_{ij}\delta_{kl}\Big) \frac{\partial}{\partial \J_2}
+ \frac{3}{5}\left(\delta_{ik}\delta_{jl}+\delta_{il}\delta_{jk}-\frac{2}{3}\delta_{ij}\delta_{kl}\right)\nonumber \\
&\qquad \times \J_2 \frac{\partial^2}{\partial \J_2^2} 
+\frac{3}{5}\left(\delta_{ik}\delta_{jl}+\delta_{il}\delta_{jk}-\frac{2}{3}\delta_{ij}\delta_{kl}\right) \Q_2
\frac{\partial^2}{\partial \J_2 \partial \Q_2} \nonumber \\
& \quad + \frac{3}{20} \left(\delta_{ik}\delta_{jl}+\delta_{il}\delta_{jk}
-\frac{2}{3}\delta_{ij}\delta_{kl}\right) \K_2 \frac{\partial^2}{\partial \Q_2^2}\bigg] p(\vy) \;,
\end{align}
and
\begin{align}
\left\langle\frac{\partial^2}{\partial\zeta_{ij}\partial \K_{kl}}\right\rangle_\Omega p(\vy) &= 
\bigg[-\delta_{ij}\delta_{kl} \frac{\partial^2}{\partial\nu\partial \J_1}
+ \frac{1}{2}\Big(3\delta_{ik}\delta_{jl}-\delta_{ij}\delta_{kl}\Big) \frac{\partial}{\partial \Q_2} 
+\frac{3}{5}\bigg(\delta_{ik}\delta_{jl}+\delta_{il}\delta_{jk} \nonumber \\
&\quad -\frac{2}{3}\delta_{ij}\delta_{kl}\bigg) \Q_2 \frac{\partial^2}{\partial \J_2 \partial \K_2}
+\frac{3}{10} \left(\delta_{ik}\delta_{jl}+\delta_{il}\delta_{jk}-\frac{2}{3}\delta_{ij}\delta_{kl}\right)
\bigg(\J_2 \frac{\partial^2}{\partial \J_2\partial \Q_2} \nonumber \\ 
&\quad +\K_2 \frac{\partial^2}{\partial \K_2\partial \Q_2}\bigg) 
+\frac{3}{20}\left(\delta_{ik}\delta_{jl}+\delta_{il}\delta_{jk}-\frac{2}{3}\delta_{ij}\delta_{kl}\right)
\Q_2 \frac{\partial^2}{\partial \Q_2^2}\bigg] p(\vy)
\end{align}
We eventually arrive at Eq.~(\ref{eq:DDav}) upon combining the above results.

\section{Non-Gaussian bias in microscopic Lagrangian halo approaches}
\label{app:ngbias}

Here, we derive how \refeq{bphi} is obtained starting from a general microscopic Lagrangian bias expansion Eq.~(\ref{eq:Fdpkeff}).
Since primordial non-Gaussianity is imprinted in the initial conditions, we can immediately insert the primordial 3-point function
into the Lagrangian expansion. For sake of clarify, let us restore the explicit time dependence of the linear density field:
$\delta^{(1)}\equiv \delta^{(1)}(\vk,\tau)$.

For a non-vanishing primordial 3-point function, the Lagrangian-space power spectrum of halos includes a term of the form
\begin{equation}
  \label{eq:phhL}
  P_{hh}^L(k) \supset 2\cdot\frac{1}{2}c_1^L(k) \int_{\vk_1}\!\int_{\vk_2}\! c_2^L(\vk_1,\vk_2)
  \big\langle\delta^{(1)}(\vk_1,\tau)\delta^{(1)}(\vk_2,\tau)\delta^{(1)}(\vk,\tau)\big\rangle \;.
\end{equation}
Expressing the linear density field $\delta^{(1)}$ in terms of the curvature perturbation $\Phi$ deep in matter domination, that is,
\begin{equation}
\delta^{(1)}(\vk,\tau) = \frac{2 k^2 T(k) D(\tau)}{3 \Omega_m H_0^2} \Phi(\vk) \equiv \mathcal{M}(k,\tau) \Phi(\vk) 
\end{equation}
where $\tau$ is the conformal time, we have
\begin{align}
  \big\langle\delta^{(1)}(\vk_1,\tau)\delta^{(1)}(\vk_2,\tau)\delta^{(1)}(\vk_3,\tau)\big\rangle &= (2\pi)^3 \delta_D(\vk_{123})
  \mathcal{M}(k_1,\tau) \mathcal{M}(k_2,\tau) \mathcal{M}(k_3,\tau) \nonumber \\
  &\qquad \times B_\Phi(k_1,k_2,k_3) \;.
  \label{eq:3pt}
\end{align}
Here, $B_\Phi(k_1,k_2,k_3)$ is the primordial bispectrum of curvature perturbations.
For local-type quadratic primordial non-Gaussianity, we have
\begin{equation}
  \label{eq:bloc}
B_\Phi(k_1,k_2,k_3) = 2 f_\text{NL} \Big[ P_\Phi(k_1) P_\Phi(k_2) + \mbox{2 cyc.}\Big] \;.
\end{equation}
Substituting Eq.~(\ref{eq:3pt}) into Eq.~(\ref{eq:phhL}), we obtain 
\begin{equation}
  P_{hh}^L(k) \supset \mathcal{M}(k,\tau) c_1^L(k) \int_{\vk_1}\!c_2^L(\vk_1,\vk-\vk_1)\, \mathcal{M}(k_1,\tau) \mathcal{M}(|\vk-\vk_1|,\tau)
  B_\Phi(k_1,|\vk-\vk_1|,k) \;.
\end{equation}
Taking the limit $k\to 0$ and assuming local-type non-Gaussianity [Eq.~(\ref{eq:bloc})], the non-Gaussian contribution reduces to
\begin{equation}
  P_{hh}^L(k) \supset 2 c_1^L(k) \left(\frac{2 f_\text{NL}b_\phi}{\mathcal{M}(k)}\right)\plin(k,\tau) \;,
\end{equation}
where $\plin(k,\tau)$ is the linear density power spectrum at the collapse epoch, and the amplitude $b_\phi$ of the non-Gaussian bias is
given by Eq.~(\ref{eq:bphi}). Note that Eq.~(\ref{eq:bphi}) holds regardless of the precise shape of the primordial curvature bispectrum.

\bibliographystyle{JHEP}
\bibliography{references}

\label{lastpage}

\end{document}